\documentclass[twocolumn]{emulateapj}
\newcommand{\nd}{\multicolumn{1}{c}{$\dots$}}

\newcommand{\msa}{mag/\sq\arcsec}
\newcommand\T{\rule{0pt}{3ex}}
\newcommand\B{\rule[-1.5ex]{0pt}{0pt}}

\slugcomment{Accepted for publication in the Astronomical Journal}
\shorttitle{Blending of Cepheids in M33}
\shortauthors{Chavez, Macri \& Pellerin}

\begin{document}
\title{Blending of Cepheids in M33}

\author{Joy M. Chavez$^1$, Lucas M. Macri \& Anne Pellerin$^2$}
\affil{George P. and Cynthia Woods Mitchell Institute in Fundamental Physics and Astronomy,\\
Department of Physics and Astronomy, Texas A\&M University,
4242 TAMU, College Station, TX 77843-4242}

\footnotetext[1]{Current address: Gemini Observatory, Northern Operations Center, Hilo, HI, 96720. jchavez@gemini.edu}

\footnotetext[2]{Current address: Department of Physics, Mount Allison University, Sackville, NB E4L 1E6, Canada.}

\begin{abstract}
A precise and accurate determination of the Hubble constant based on Cepheid variables requires proper characterization of many sources of systematic error. One of these is stellar blending, which biases the measured fluxes of Cepheids and the resulting distance estimates. We study the blending of 149 Cepheid variables in M33 by matching archival {\it Hubble Space Telescope} data with images obtained at the WIYN 3.5-m telescope, which differ by a factor of 10 in angular resolution.

We find that $55\pm4$\% of the Cepheids have no detectable nearby companions that could bias the WIYN $V$-band photometry, while the fraction of Cepheids affected below the 10\% level is $73\pm4$\%. The corresponding values for the $I$ band are $60\pm4$\% and $72\pm4$\%, respectively. We find no statistically significant difference in blending statistics as a function of period or surface brightness. Additionally, we report all the detected companions within 2$\arcsec$ of the Cepheids (equivalent to 9 pc at the distance of M33) which may be used to derive empirical blending corrections for Cepheids at larger distances. 
\end{abstract}

\keywords{Cepheid Variables; Galaxies: Individual (M33)}

\section{Introduction}
An accurate and precise measurement of the Hubble constant at the few-percent level imposes significant constraints on the equation of state of dark energy and other cosmologically relevant parameters \citep{komatsu11}. The next generation of surveys aimed at improving our understanding of dark energy will benefit from an even tighter constraint on $H_0$ \citep{weinberg12} than the present bounds of 3.4\% \citep{riess11}.

Cosmological applications of the Extragalactic Distance Scale \citep{freedman10} primarily rely on the Period-Luminosity relation of Cepheid variables \citep[hereafter the ``Leavitt law'',][]{leavitt12} as the primary distance indicator. The upcoming {\it Gaia} mission \citep{prusti11} is expected to deliver a sub-percent calibration of the Leavitt law in the Milky Way \citep{windmark11}, which could in turn enable a 1\% measurement of $H_0$ if all sources of systematic error are properly accounted for.

One of these sources of systematic error occurs when two or more neighboring (but not necessarily physically associated) stars fall within the same resolution element of an instrument and cannot be fit with separate point-spread functions (PSFs). This effect is commonly referred to as {\it blending} and it is different from {\it crowding} or confusion noise, which results in improper PSF fitting and/or inaccurate background subtraction due to a very high stellar density. An extreme example of blending in the absence of crowding is a Cepheid in a binary system located in a low-surface brightness environment. Blending will bias the measured flux of a Cepheid towards larger values, shifting the Leavitt law to brighter magnitudes and leading to systematically shorter distances and larger values of $H_0$. Extreme blends can be readily identified by their effects on Cepheid colors and/or amplitude ratios and such tests are routinely carried out \citep{pellerin11,scowcroft09,macri06}. However, low-level blends are unlikely to be identified by such cuts and may affect studies of the metallicity dependence of the Leavitt law (another source of systematic uncertainty) since they could mimic the photometric changes expected from differences in chemical abundances. 

The Local Group galaxy M33 is a good testbed for studies of Cepheid systematics thanks to its relative proximity \citep[$D=895-965$~kpc,][]{bonanos06,pellerin11}, moderate inclination angle \citep[$i=55^{\circ}$,][]{ho97} and recent episodes of star formation which have resulted in large numbers of Cepheids throughout its disk \citep{hartman06,pellerin11}. \citet{scowcroft09} used M33 Cepheids to study the ``metallicity effect'' of the Leavitt law, motivated by the large abundance gradient inferred from \ion{H}{2} regions \citep{zaritsky94,magrini07,magrini10}. However other studies \citep{urbaneja05,bresolin10,bresolin11} have determined a much shallower abundance gradient, which would make the metallicity effect considerably harder to measure.

The disk of M33 has been extensively imaged by the Hubble Space Telescope (HST) using the Wide-Field and Planetary Camera 2 (WFPC2) and the Advanced Camera for Surveys (ACS). The angular resolution of HST at optical wavelengths is 10-15 times better than the seeing at a good site on the surface of the Earth. Thus, a comparison of HST and ground-based images of the same Cepheids in M33 can yield useful insights into the nature of blending for more distant galaxies observed only with Hubble. 

\begin{figure*}[t]
\centering
 \includegraphics[angle=0,width=\textwidth]{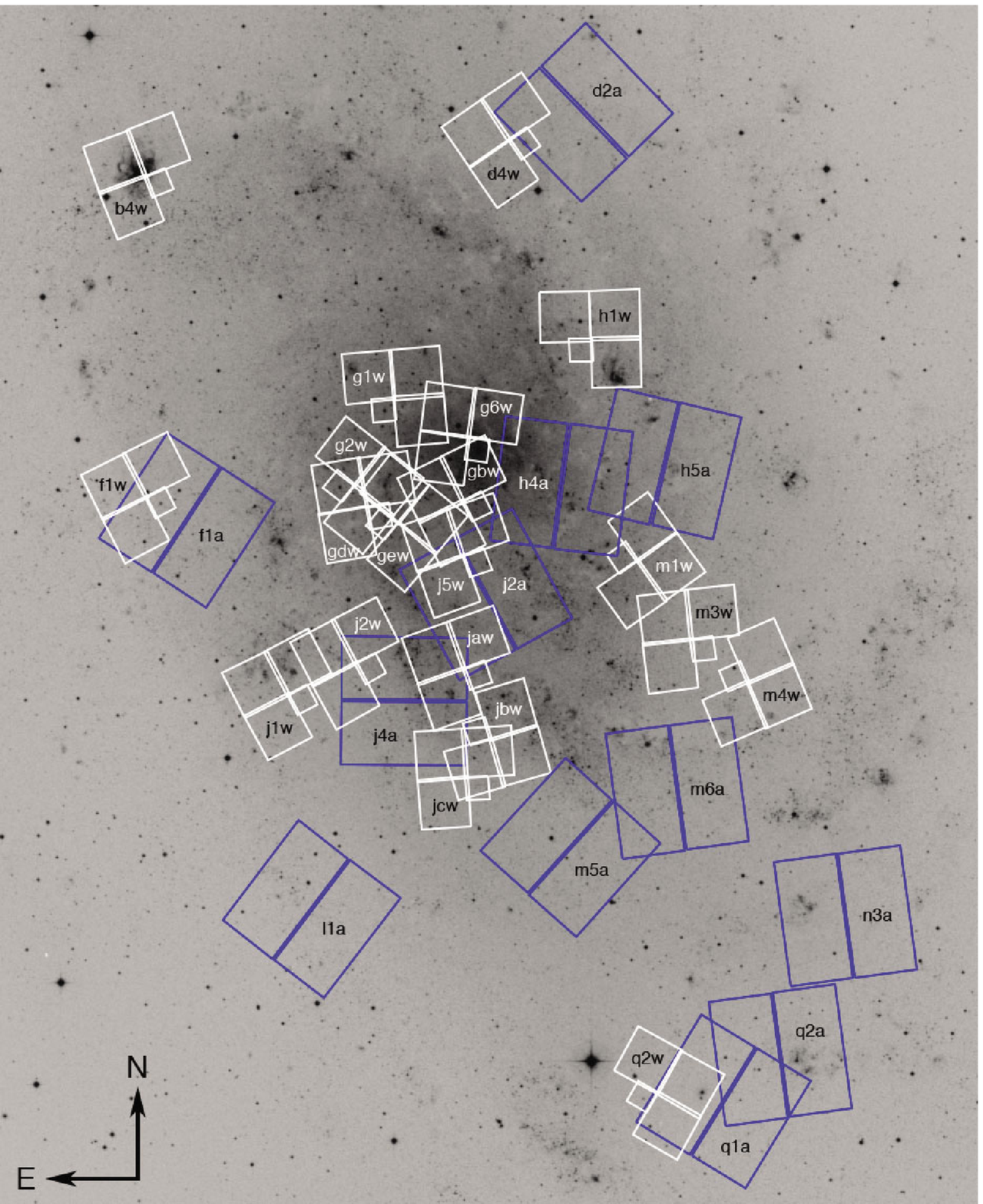}
 \caption{Footprints of the HST fields used in this study overlayed on a DPOSS-II image of M33.  The blue rectangles are from ACS, and the white boxes are from WFPC2.  The field label names end in 'a' for ACS, and 'w' for WFPC2.}
 \label{fig:footprints}
\end{figure*}

\begin{deluxetable*}{llllcrcrrcl}
\tabletypesize{\footnotesize}
\tablewidth{0pc}
\tablecaption{HST observations of M33 used in this study \label{tb:fields}}
\tablehead{\colhead{Field} & \colhead{RA}    & \colhead{Dec}   & \colhead{Camera} & \multicolumn{2}{c}{Filters} & \multicolumn{2}{c}{Exp. time (s)}  &\colhead{Prop.} & \colhead{N}   & \multicolumn{1}{l}{Comments}   \\ 
           \colhead{Name } & \multicolumn{2}{c}{(J2000, deg)}  & \colhead{}      & \colhead{\#1} &\colhead{\#2} & \colhead{\#1} &\colhead{\#2}       &\colhead{\#}   & \colhead{Ceph.} & \colhead{}} 
\startdata
b4w & 23.6384 & 30.7818 & WFPC2 & 555 & 814 & 400   & 400   &  5237 &  3 & 16 in P07 \\
d2a & 23.4066 & 30.8111 & ACS   & 606 & 814 & 10414 & 20828 &  9873 &  7 & U49 in S06\\
d4w & 23.4519 & 30.7977 & WFPC2 & 555 & 814 & 4800  & 5200  &  5914 &  5 & U49 in K02 \\
f1a & 23.6108 & 30.6291 & ACS   & 606 & 814 & 10414 & 20828 &  9873 &  6 & M9 in S06\\
f1w & 23.6388 & 30.6393 & WFPC2 & 555 & 814 & 4800  & 5200  &  5914 &  1 & 2 in C01\\
g1w & 23.5049 & 30.6843 & WFPC2 & 555 & 814 & 4800  & 5200  &  5914 &  3 & R14 in K02 \\
g2w & 23.5185 & 30.6446 & WFPC2 & 555 & 814 & 4800  & 5200  &  5914 & 13 & 10 in C01\\
g6w & 23.4670 & 30.6671 & WFPC2 & 555 & 814 & 1600  & 1200  &  5464 &  2 & \\
gbw & 23.4753 & 30.6373 & WFPC2 & 555 & 814 & 2240  & 2120  &  6640 &  2 & WW1 in C99\\
gdw & 23.5209 & 30.6349 & WFPC2 & 606 & 814 & 340   & 400   &  8059 &  2 & \\
gew & 23.4963 & 30.6280 & WFPC2 & 606 & 814 & 1600  & 360   &  8059 &  5 & also prop. \#8805\\
h1w & 23.4030 & 30.7095 & WFPC2 & 606 & 814 & 8800  & 17600 &  9873 &  4 & \\
h4a & 23.4178 & 30.6439 & ACS   & 606 & 814 & 2160  & 2160  & 10190 &  6 & F1 in S09\\
h5a & 23.3644 & 30.6539 & ACS   & 606 & 814 & 2400  & 2500  & 10190 &  7 & F2 in S09\\
j1w & 23.5654 & 30.5517 & WFPC2 & 606 & 814 & 1600  & 360   &  8059 &  2 & also prop. \#8805\\
j2a & 23.4581 & 30.5973 & ACS   & 606 & 814 & 6380  & 6624  & 10190 &  8 & innermost in W09; D1 in S09\\
j2w & 23.5308 & 30.5654 & WFPC2 & 606 & 814 & 300   & 700   &  8059 &  2 & \\
j4a & 23.5010 & 30.5487 & ACS   & 606 & 814 & 2400  & 2500  & 10190 &  5 & F3 in S09\\
j5w & 23.4733 & 30.6000 & WFPC2 & 555 & 814 & 2240  & 2120  &  6640 &  2 & 15 in C01\\
jaw & 23.4735 & 30.5632 & WFPC2 & 555 & 814 & 2240  & 2080  &  6640 &  2 & 17 in C01\\
jbw & 23.4523 & 30.5314 & WFPC2 & 555 & 814 & 2240  & 2080  &  6640 &  4 & \\
jcw & 23.4697 & 30.5199 & WFPC2 & 555 & 814 & 2240  & 2080  &  6640 &  1 & \\
l1a & 23.5497 & 30.4560 & ACS   & 606 & 814 & 2160  & 2160  & 10190 &  1 & F5 in S09\\
m1w & 23.3728 & 30.6106 & WFPC2 & 555 & 814 & 520   & 460   &  6431 & 10 & \\
m3w & 23.3535 & 30.5758 & WFPC2 & 606 & 814 & 1600  & 360   &  8059 & 10 & also prop. \#8805\\
m4w & 23.3179 & 30.5566 & WFPC2 & 606 & 814 & 240   & 400   &  8059 &  4 & \\
m5a & 23.4161 & 30.4834 & ACS   & 606 & 814 & 21260 & 26420 & 10190 &  4 & D2 in S09, 26 in C01\\
m6a & 23.3592 & 30.5109 & ACS   & 606 & 814 & 2160  & 2160  & 10190 & 15 & F4 in S09\\
n3a & 23.2727 & 30.4538 & ACS   & 606 & 814 & 2400  & 2500  & 10190 &  4 & F6 in S09 \\
q1a & 23.3348 & 30.3706 & ACS   & 606 & 814 & 21260 & 26420 & 10190 &  6 & D3 in S09 \\
q2a & 23.3068 & 30.3919 & ACS   & 606 & 814 & 2160  & 2160  & 10190 &  3 & 3rd outermost in W09; F7 in S09\\
q2w & 23.3645 & 30.3742 & WFPC2 & 555 & 814 & 4800  & 5200  &  5914 &  1 & 9 in C01; C38 in K02
\enddata
\tablecomments{C99: \citet{chandar99}; C01: \citet{chandar01}; K02: \citet{kim02}; P07: \citet{park07}; S06: \citet{sarajedini06}; S09: \citet{sanroman09}; W09: \citet{williams09}}
\end{deluxetable*}

Previous studies of the influence of blends on the Cepheid Distance Scale, based on comparisons between ground-based and HST images of nearby galaxies were carried out by \citet{mochejska00} in M31, by \citet{mochejska01} in M33, and by \citet{bresolin05} in NGC$\,$300. In the case of M33, \citet{mochejska01} used HST/WFPC2 images and the Cepheid sample of the DIRECT survey \citep{macri01}. During the intervening decade there have been numerous additional HST observations of M33 using both WFPC2 and the Advanced Camera for Surveys (ACS), which enable us to study more Cepheids than \citet{mochejska01} and, in the case of ACS, with greater depth and finer pixel scale. Furthermore, we rely on a new synoptic survey of M33 \citep{pellerin11} carried out at the WIYN 3.5-m telescope with more Cepheids and better angular resolution than the DIRECT catalog.

\citet{pellerin11} carried out extensive simulations based on the M33 ACS images to quantify the photometric bias due to crowding in their ground-based photometry. Considering the range of magnitudes and surface brightnesses spanned by the M33 Cepheid sample, they found that crowding bias increased as a function of magnitude but did not exhibit a dependence on surface brightness. Our paper complements their study by quantifying the photometric bias due to blending for Cepheids in M33.

We describe in \S\ref{sec:data} the data used in this paper and the photometry we measured; \S\ref{sec:method} describes the method used to quantify the level of blending; we discuss the results in \S\ref{sec:results} and compare them to previous work in \S\ref{sec:comp}. Our concluding remarks and suggestions for future work can be found in \S\ref{sec:conclusions}.

\section{Data and Analysis}\label{sec:data}

We based our analysis on the Cepheid sample published by the M33 Synoptic Stellar Survey \citep{pellerin11}. We identify these variables in HST images and search for companions unresolved in the ground-based data. We calculate blending statistics based on these companions.

\subsection{Cepheid Sample}
Our analysis is based on the sample of Cepheids listed in Table~3 of \citet{pellerin11}. The ground-based observations and analysis are described in detail in that publication, which we briefly summarize here. Data from the DIRECT survey of M33 \citep{macri01} were combined with new images obtained at the 3.5-m Wisconsin-Indiana-Yale-NOAO (WIYN) telescope with the Mini-Mosaic (MiniMo) camera to detect 563 Cepheids ranging in period from 2 to 110 days. The typical FWHM of the WIYN images was $0\farcs75$, sampled at a plate scale of $0\farcs28/$pix. The photometry and astrometry were calibrated using the catalogs of \citet{massey06}. 

\subsection{HST Data}
We queried the Hubble Legacy Archive (HLA) and the Mikulski Archive for Space Telescopes (MAST)\footnote{The HLA and MAST are operated by the Space Telescope Science Institute (STScI).} for HST images of M33 obtained with either WFPC2 or ACS which had overlap with the WIYN images of \citet{pellerin11}. We selected observations with multiple exposures to allow for cosmic-ray removal. We also required a minimum of 100\,s of total exposure time, to ensure a depth that would enable the detection of faint companions around the Cepheids. We further restricted our study to fields that were imaged in $V$ (HST filters F555W or F606W) and $I$ (HST filter F814W).

The HST fields contained 149 ($\sim 25$\%) of the Cepheids listed in \citet{pellerin11}. The locations of these fields are shown in Figure \ref{fig:footprints} and listed in Table \ref{tb:fields}. The table also contains references to previously-published analyses of the data. Except for two ACS fields, all images were acquired on a single epoch and we therefore only have imaging of the Cepheids at a random phase within their pulsation cycle. 

The ACS images were reprocessed through the MAST On-The-Fly-Recalibration pipeline to apply the most up-to-date calibrations, while the WFPC2 images had already been reprocessed using the final set of calibration frames in mid 2009 by STScI \citep{gonzaga10}. We downloaded the reprocessed images and used MultiDrizzle \citep{koekemoer02} to remove cosmic rays, correct for geometric distortions in the cameras, and co-add multiple observations into master images.

\subsection{Photometry and Cepheid Search \label{sec:photometry}}

We performed point-spread function (PSF) photometry using DAOPHOT and ALLSTAR \citep{stetson87}. We derived model PSFs using grids of artificial stars created with TinyTim \citep{krist04} for the appropriate bandpasses, cameras and CCDs. We ran the FIND algorithm twice on each image, removing all stars found on the first iteration before proceeding to the second one. This increased the detection efficiency of faint stars, such as possible companions of a Cepheid. ALLSTAR was run one final time on the merged star list. Based on the observed luminosity functions, the photometry is complete to $V\!\sim\!25.5, I\!\sim\!24.7$ and $V\!\sim\!24.3, I\!\sim\!23$~mag for ACS and WFPC2, respectively.

\vspace{2pt}

Instrumental magnitudes were converted to the HST VEGAMAG system using the equations listed in Appendix D of \citet{sirianni05} and the coefficients listed in Table 10 of \citet{sirianni05} and Table 2 of \citet{dolphin09} for ACS and WFPC2, respectively. 

\vspace{2pt}
 
\begin{figure}[t]
\includegraphics[width=0.48\textwidth]{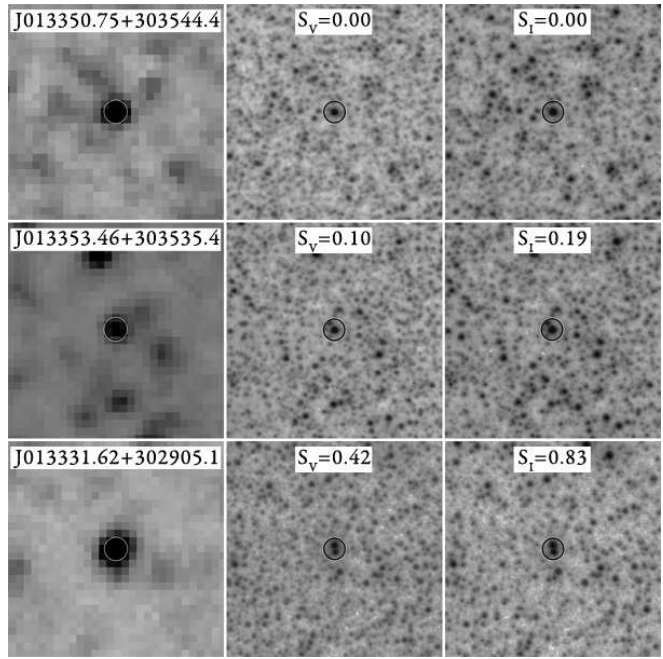}
\caption{Comparison of HST and WIYN images of three Cepheids in M33, illustrating different blending values. Left column: WIYN $V$ images; center column: HST $V$ images; right column: HST $I$ images. Panels are $8\arcsec$ on a side and the black circles are $0.75\arcsec$ in diameter, equal to the WIYN FWHM.}
\label{fig:thumbnails}
\end{figure}

Given the vastly different resolution and depth of the HST and WIYN images, the former had significantly larger stellar densities. Furthermore, the astrometric solution provided by the automated STScI pipeline is only accurate to a few arcseconds \citep{koekemoer05}. We obtained a rough initial match between HST and WIYN images using the brightest few hundred stars in common. Once the gross astrometric offset had been removed, we matched the complete star lists using DAOMATCH and DAOMASTER \citep{stetson93} and refined the astrometric solution of the HST images. Cepheids were then selected based on the coordinates tabulated by \citet{pellerin11}.  We visually inspected every Cepheid to ensure the star in the HST frame was indeed a match to the same star in WIYN image. This process helped to identify and correct a few erroneous matches where a faint star close to the Cepheid was originally identified as the variable in the HST frame. Lastly, we estimated the disk surface brightness by averaging the background flux values reported by ALLSTAR for stars within $7\arcsec$ of each Cepheid. 

\section{Blending calculation}\label{sec:method}
We quantify the level of blending following the prescription of \citet{mochejska00},

\vspace{3pt}

\begin{equation} \label{eq:blend}
S_{F}=\sum(f_i) / f_C
\end{equation}

\vspace{3pt}

\noindent{where $S_F$ is the total flux contribution from the companions relative to the Cepheid in filter $F$, $f_i$ is the flux of an individual companion star located within the critical radius and $f_C$ is the flux of the Cepheid.}

\vspace{3pt}

We calculated the values of $S$ separately for $V$ and $I$, using a critical radius of $0\farcs375$ which is the average value of the half-width at half-maximum (HWHM) of the WIYN PSF. We only include companions that contribute 4\% or more of the flux of a Cepheid in order to provide a conservative estimate of the blending value. This cut-off was adopted by \citet{mochejska00}, although \citet{mochejska01} raised it to 6\%. In practice, stars with $f_i \sim 0.05 f_C$ (or $\Delta$mag$\sim 3.25$) are near the completeness limit of the ACS images relative to the faintest, shortest-period Cepheids, which have $V\sim22.5, I\sim21.5$~mag. 

\vspace{3pt}
 
In the case of Cepheids present in both ACS and WFPC2 images, we calculated blending values using the ACS data given its finer spatial sampling and increased depth. In the case of Cepheids present in multiple fields obtained with the same camera, we gave preference to the image with the deepest exposure time. If the exposures were of similar depth, we averaged the Cepheid magnitudes and the $S$ values.

\vspace{3pt}
 
Figure~\ref{fig:thumbnails} shows a comparison of HST and WIYN images for three Cepheids with different values of $S$. Each panel is $8\arcsec$ on a side, centered on a Cepheid. Circles with radii of $0\farcs375$ (typical WIYN HWHM) are drawn around the variables. The Cepheids were chosen to show the range of blending values, from $S_F=0$ (top row) to $S_F\sim0.6$ (bottom row).  The left column shows the WIYN $V$ images, while the center and right columns show the $V$ and $I$ HST images.

\vspace{2pt}
  
The photometry and blending values are listed in Table \ref{tb:cephblend}. For each Cepheid, we list the ID and period from \citet{pellerin11}, the $V$ magnitude and its uncertainty, the value of $S_V$ and its uncertainty, and the corresponding information for the $I$ band. Additionally, we tabulate the $V$ and $I$ surface brightness values and the designations of the WIYN and HST fields where each Cepheid is located. The uncertainties in $S_F$ values are calculated by propagating the reported ALLSTAR photometric uncertainties through Eqn.~\ref{eq:blend}. HST field codes are based on the field name listed in the first column of Table~\ref{tb:fields}, followed by a letter to identify the camera (`a' for ACS, `w' for WFPC2). 

\vspace{2pt}

\begin{figure}[t]
\includegraphics[width=0.48\textwidth]{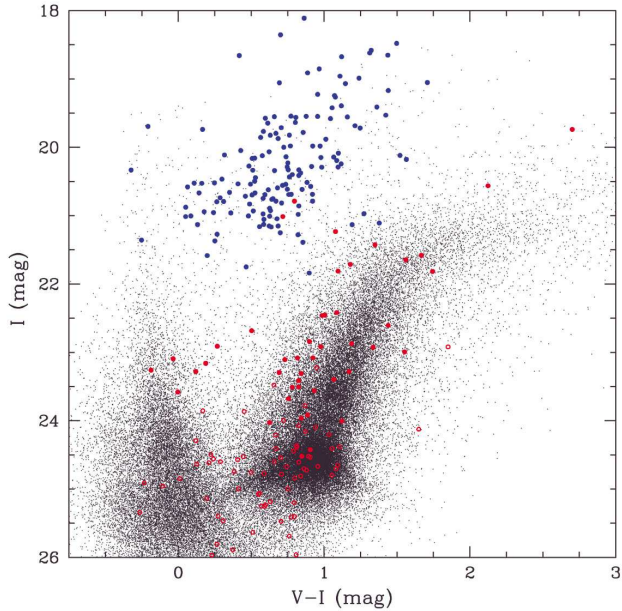}
\caption{Color-magnitude diagram of M33 Cepheids (in blue) and companions within $0\farcs 375$ (in red), contributing more (filled) or less (open) than 4\% of the Cepheid flux. Black dots are used to plot 3.5\% of the stars detected in the ACS frames with $I<26$~mag.}
\label{fig:cmd}
\end{figure}

Figure~\ref{fig:cmd} shows a color-magnitude diagram of the Cepheids and all companions located within the critical radius. As a reference, we also plot 3.5\% of all the stars with $I<26$~mag detected in the $V$ and $I$ ACS frames. The companions span a broad range of colors and magnitudes, but most are associated with the red giant branch and the red clump. These findings are not directly applicable to all Cepheid hosts, since different star-formation histories will alter the relative contributions of the upper main-sequence and the red giant branch.

\vspace{2pt}

We used the HST star lists obtained in \S\ref{sec:photometry} to tabulate all companions within a 2$\arcsec$ radius of each Cepheid, presented in Table~\ref{tb:comp}. Companions are labeled using the Cepheid ID from Table~\ref{tb:cephblend} and are numbered in increasing order of radial distance from the variable. We list the x-, y- and radial distance from the Cepheid (in arcseconds), the $V$ magnitude and uncertainty, and the $I$ magnitude and uncertainty. Some companions were only detected in one band.

\vspace{2pt}

This extended dataset can be used for a variety of future studies. For example, comparisons of HST data with ground-based observations of M33 at different angular resolutions can be easily carried out by selecting the appropriate critical radius. Likewise, the sensitivity of blending values to the faint-companion cutoff limit can be explored. Lastly, suitable scaling of fluxes and angular separations can yield simulated HST images of Cepheids in similar environments out to $D\sim35$~Mpc, at which point $2\arcsec$ at the distance of M33 would be equivalent to the angular resolution of HST in the $V$ band.

\vspace{2pt}

\section{Results}\label{sec:results}
\begin{figure*}[htp]
\epsscale{0.9}\plottwo{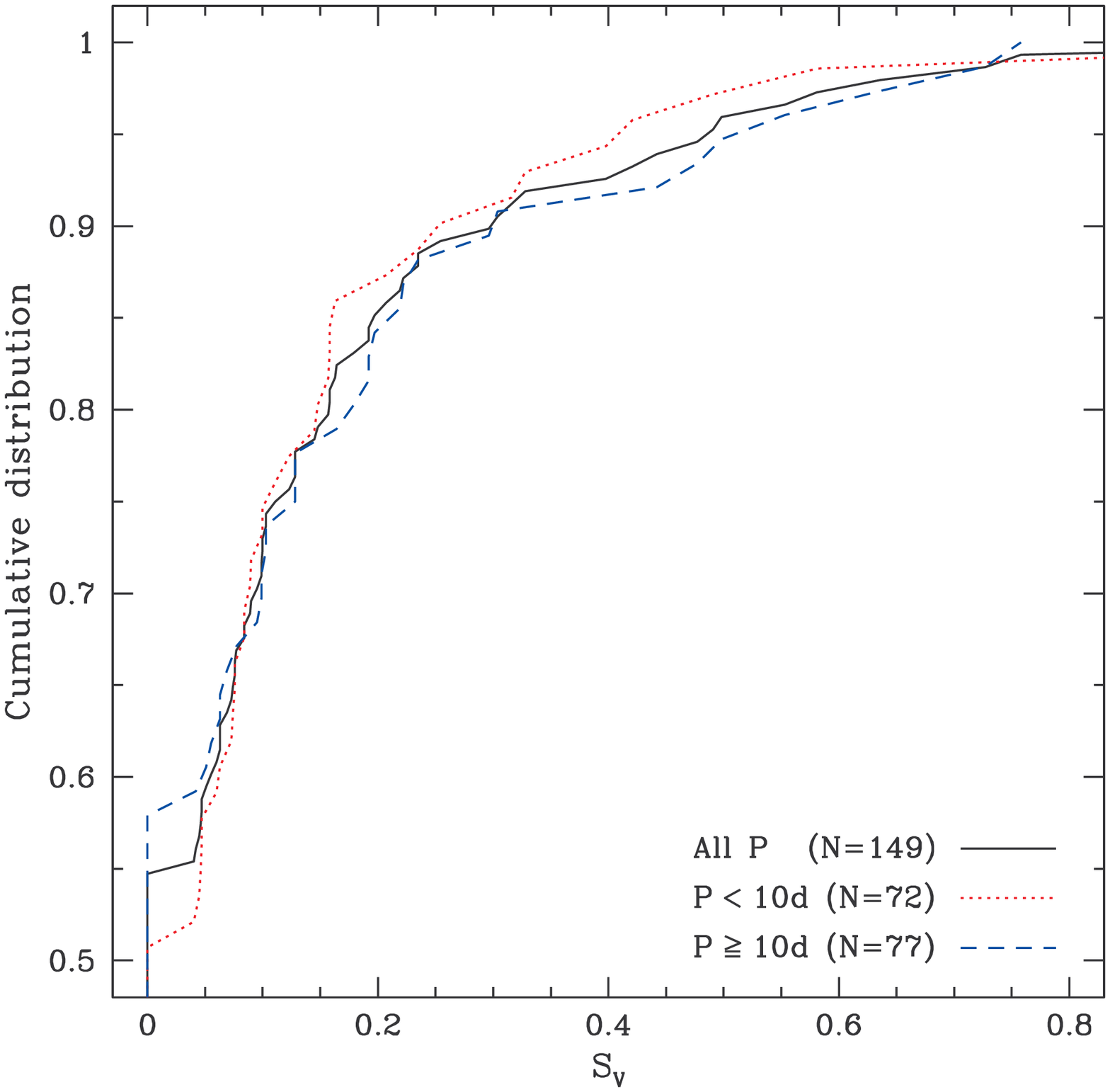}{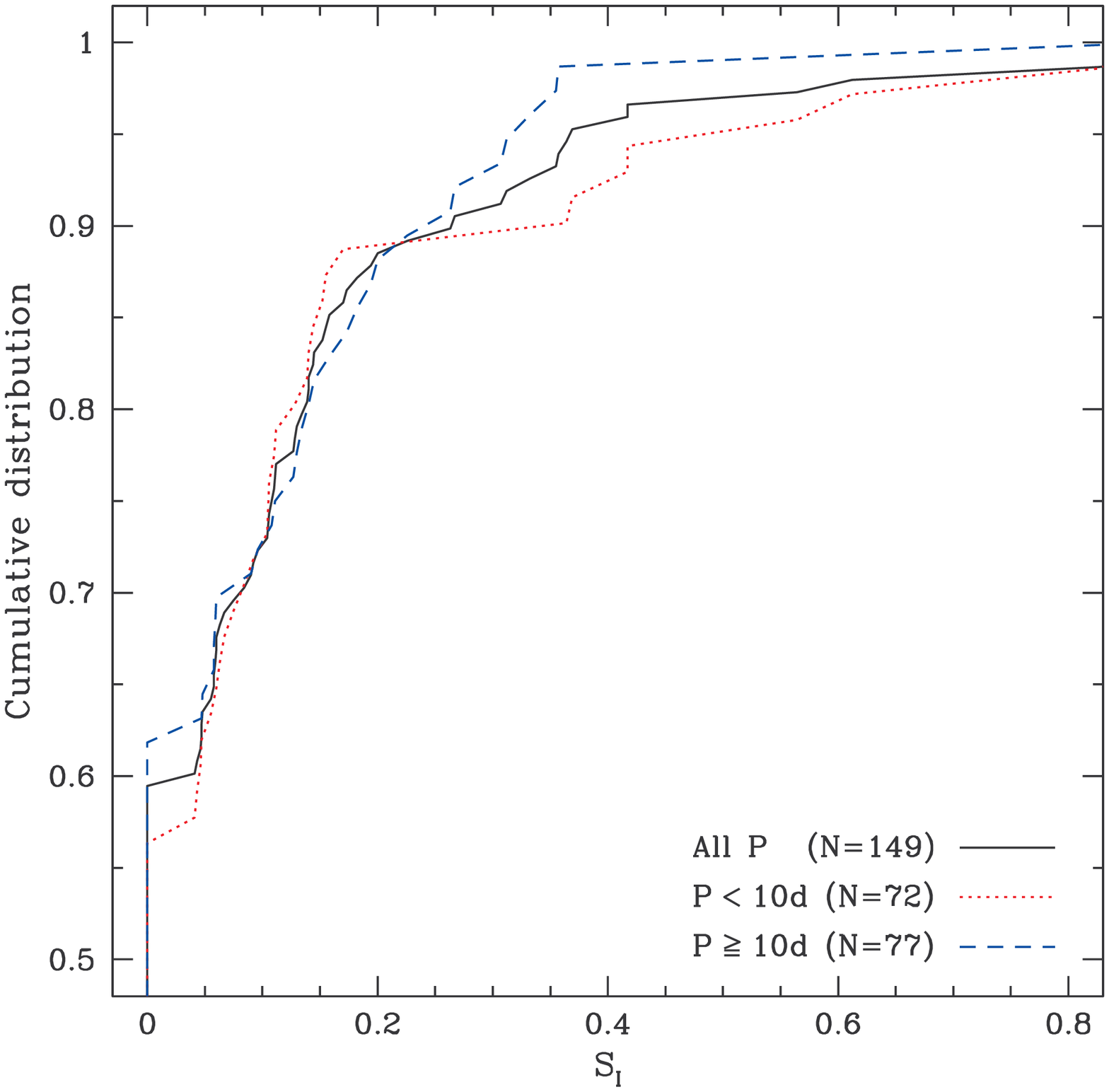}
 \caption{Cumulative distributions of the blending values in the V and I bands (right and left panels, respectively). The solid lines represent the entire Cepheid sample while the dotted and dashed lines denote the short- and long-period ($P\leq10, P>10$~d) Cepheids, respectively.}
 \label{fig:Shist}
\end{figure*}

\begin{figure*}
\epsscale{0.9}\plottwo{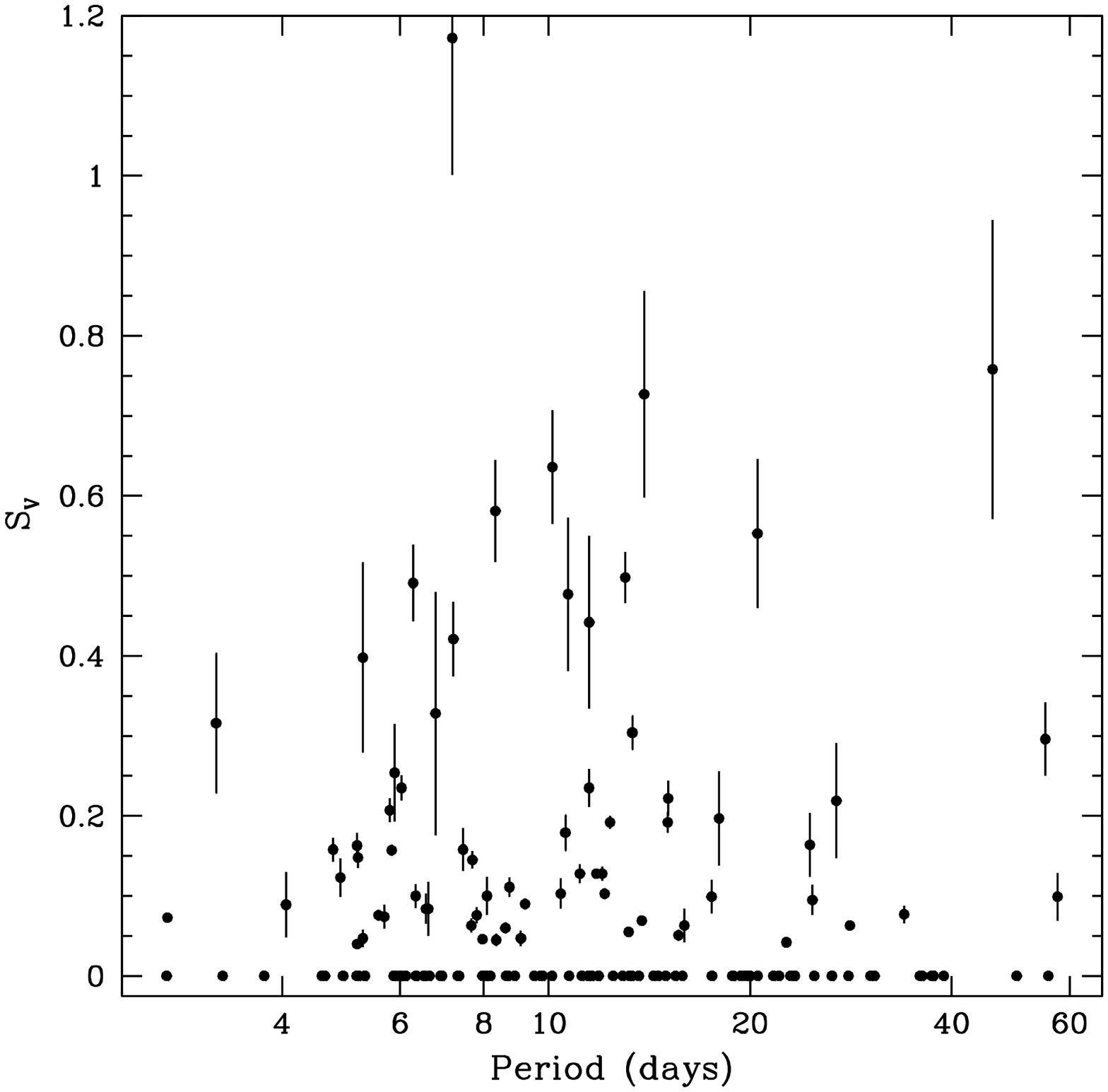}{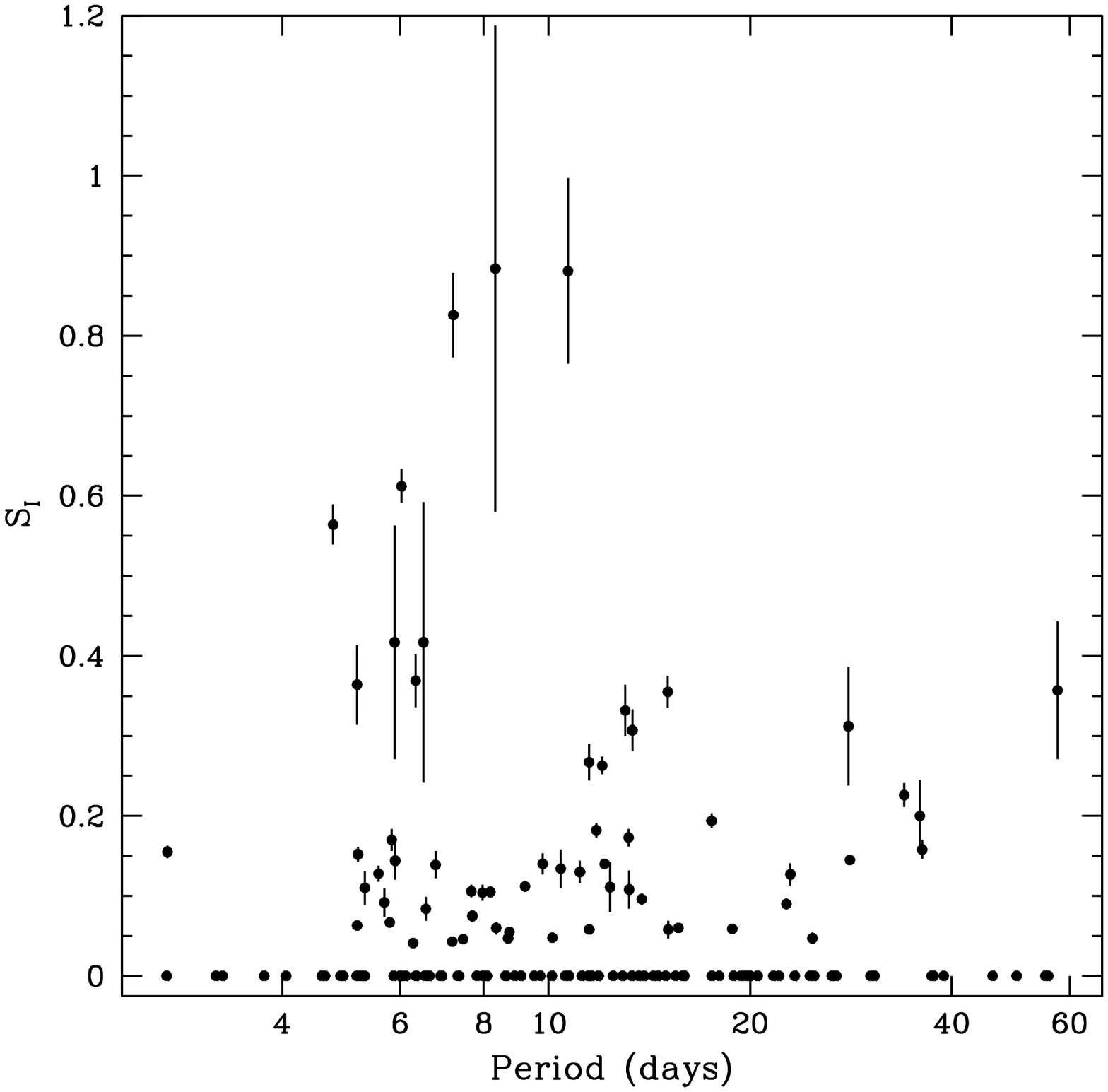}
 \caption{Blending values as a function of the period of the Cepheid.}
 \label{fig:SvP}
\end{figure*}

\begin{figure*}[htp]
\epsscale{0.9}\plottwo{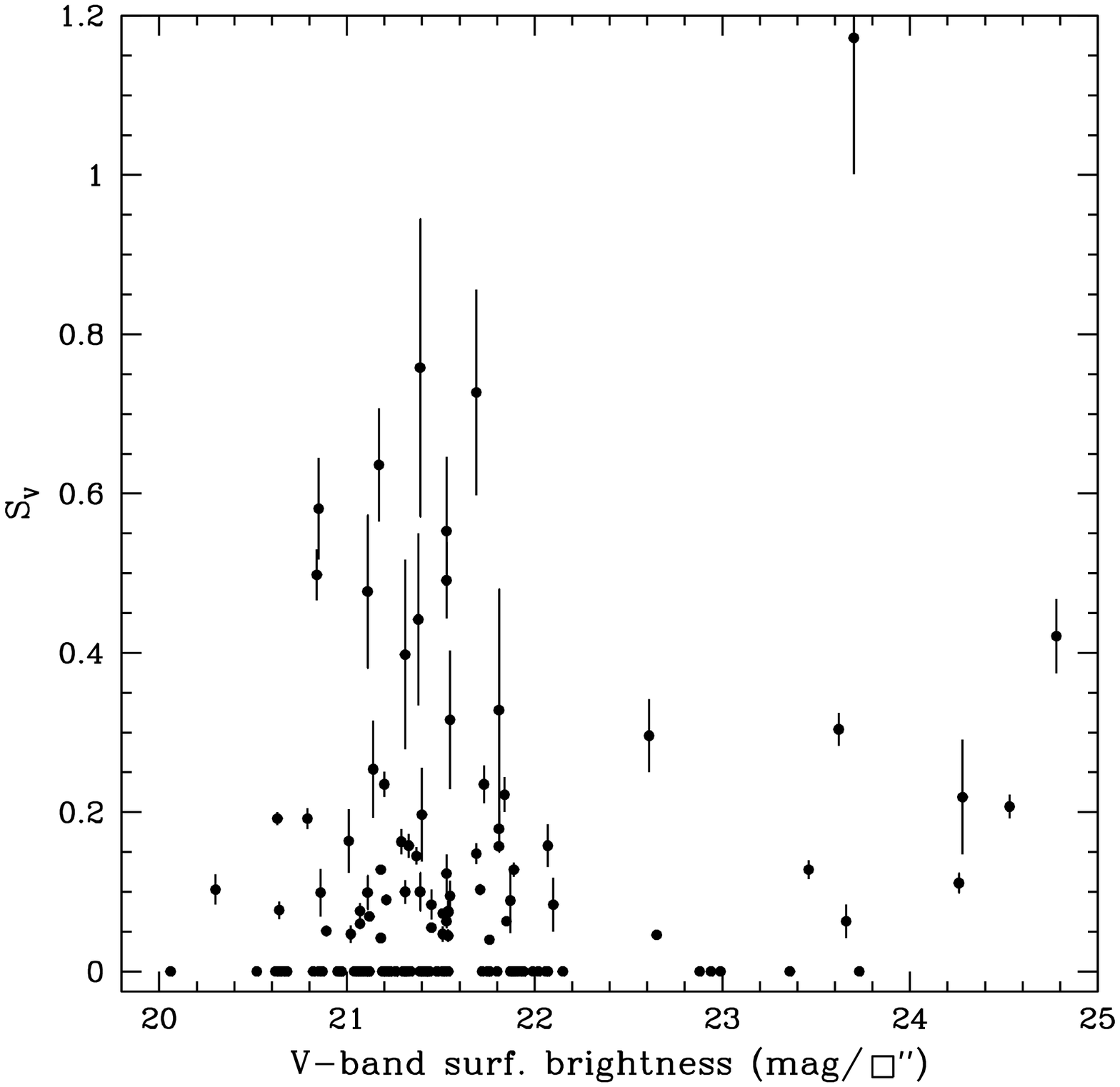}{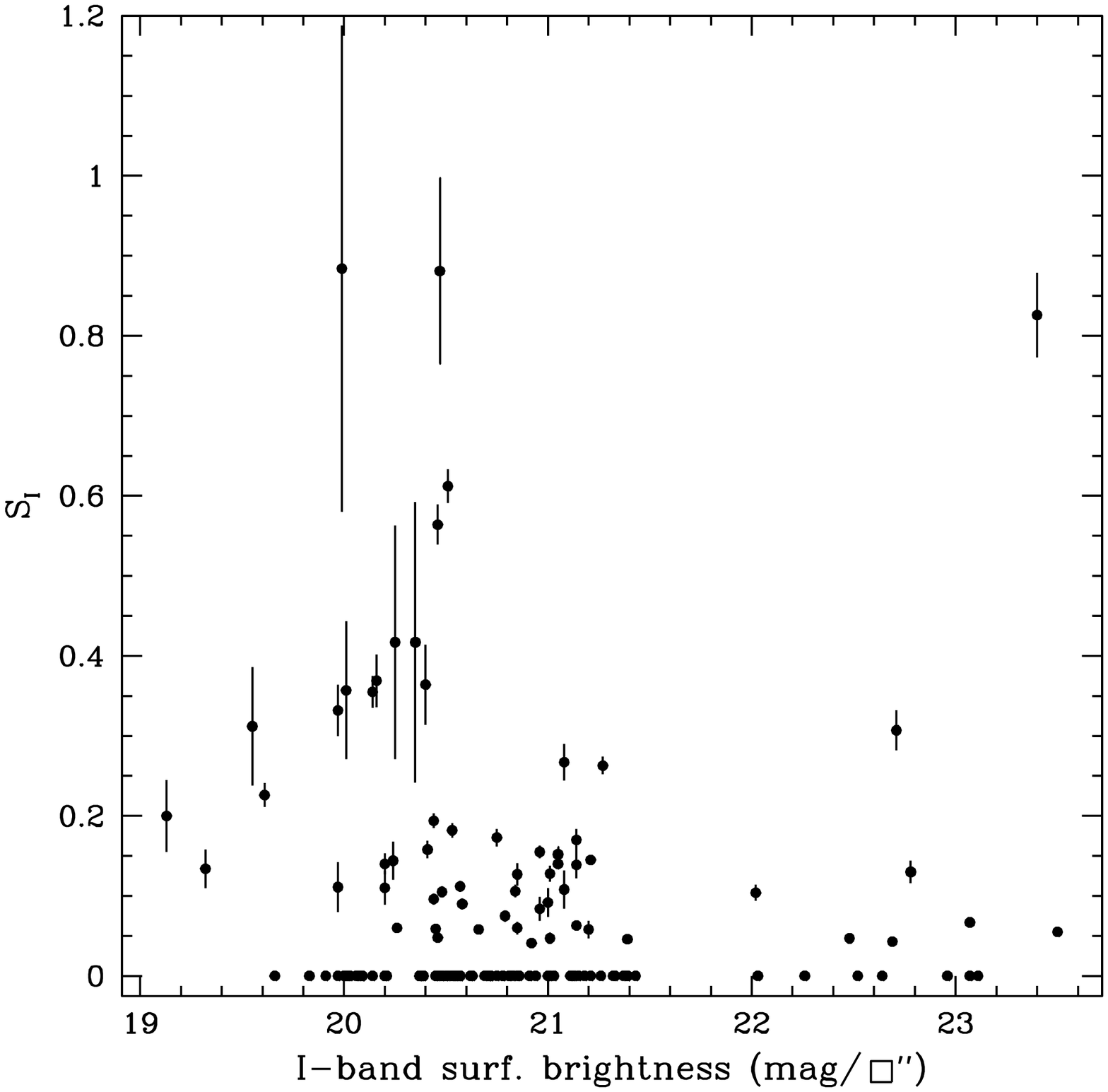}
 \caption{Blending values as a function of the sky background.}
 \label{fig:Svsky}
\end{figure*}

We find mean blending values of $S_V=0.096\pm0.015$ and $S_I=0.083\pm0.013$ and median values of zero for both bands. Figure \ref{fig:Shist} shows cumulative distributions of blending values, while Figures~\ref{fig:SvP} and \ref{fig:Svsky} show the distribution of blending values as a function of period and surface brightness, respectively. Table~\ref{tb:blendstat} lists the fractions of Cepheids which meet several blending criteria as a function of period and surface brightness. We calculated the uncertainty in each fraction using the binomial distribution approximation,

\begin{equation}
\sigma(f)=\sqrt{f (1-f) /N}
\end{equation}

\noindent{where $f$ is the fraction value and $N$ is the number of Cepheids meeting a particular set of criteria. We cross-checked the validity of this approximation by performing 100,000 bootstrap resamplings with replacement, which yielded the same uncertainties. Lastly, we tested the sensitivity to outliers in the distributions by performing the same number of jacknife resamplings, keeping 90\% of the original sample. The derived fractions remained stable at the 2\% level.}

\vspace{2pt}
The fraction of Cepheids with no blending is marginally lower ($\sim 1\sigma$) for Cepheids with $P<10$~d than for ones with $P>10$~d. Such a trend might be expected because the shorter-period, less luminous Cepheids can be affected (at a fixed flux ratio) by a larger fraction of disk stars. However, the difference vanishes when comparing the statistics of Cepheids affected at the 10\% level. There is no significant difference in the statistics of Cepheids located in areas with ``high'' or ``low'' surface brightness.

\begin{figure}[t]
\includegraphics[width=0.49\textwidth]{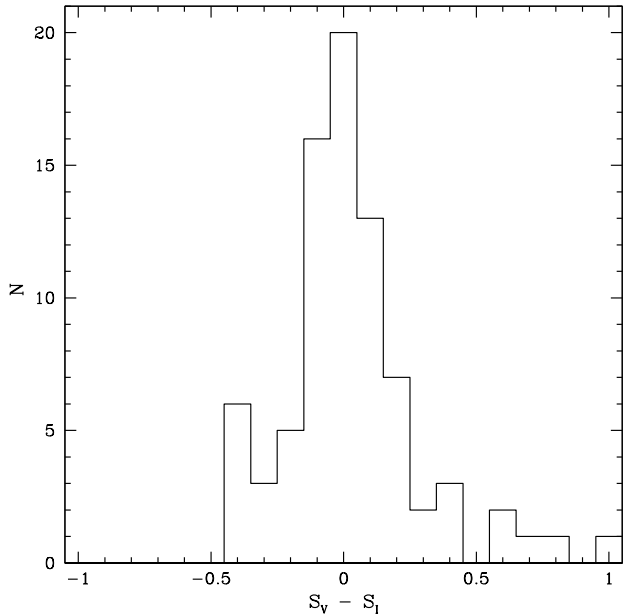}
\caption{Distribution of the ``color'' of the companions relative to their Cepheid.}
\label{fig:scolor}
\end{figure}

\vspace{2pt}
We also examined the effect of blends on the color of the Cepheids by calculating the value of $S_V-S_I$ for all Cepheids with non-zero values of either $S_V$ or $S_I$. The resulting histogram, presented in Figure \ref{fig:scolor}, shows that most blends do not appreciably change the color of the Cepheids: $\langle S_V-S_I\rangle = 0.03\pm0.27$.

\vspace{2pt}

\section{Comparison with previous work}\label{sec:comp}

\citet{mochejska01} analyzed WFPC2 images of M33 Cepheids discovered by the DIRECT project. We recalculated our blending values using the parameters adopted\ \ in\ \ that\ \ paper:\ \ a critical\ \ radius of\ \ $0\farcs75$ and 

\LongTables
\begin{deluxetable*}{lrrrrrrrrrrrrr}
\tabletypesize{\footnotesize}
\tablewidth{0pc}
\tablecaption{Cepheid properties (Abridged) \label{tb:cephblend}}
\scriptsize
\tablehead{
\colhead{ID} & \colhead{P} & \colhead{V}    & \colhead{$\sigma_V$} & \colhead{S$_V$} & \colhead{$\sigma_{S_V}$} & \colhead{I} & \colhead{$\sigma_I$} & \colhead{S$_I$} &\colhead{$\sigma_{S_I}$} &\colhead{SB$_V$} & \colhead{SB$_I$} & \multicolumn{2}{c}{Field} \\
\colhead{}   & \colhead{(d)} &\multicolumn{2}{c}{(mag)}            & \colhead{}      & \colhead{}              & \multicolumn{2}{c}{(mag)}        &\colhead{}       &\colhead{}              &\multicolumn{2}{c}{(mag/\sq\arcsec)} & \colhead{HST}   &\colhead{WIYN}}
\startdata
J013332.36+302819.8&  2.689& 21.635& 0.038 & 0.000 & 0.000 & 21.008& 0.032 & 0.000 & 0.000 & 21.92 & 21.26 & m5a & m0f\\
J013324.20+302248.9&  2.695& 21.620& 0.038 & 0.073 & 0.005 & 21.369& 0.029 & 0.155 & 0.008 & 21.51 & 20.96 & q1a & m0m\\
J013316.88+302157.9&  3.187& 22.737& 0.106 & 0.316 & 0.087 & 21.839& 0.055 & 0.000 & 0.000 & 21.55 & 21.01 & q1a & m0m\\
J013309.08+302354.6&  3.260& 22.215& 0.067 & 0.000 & 0.000 & 21.750& 0.043 & 0.000 & 0.000 & 21.94 & 21.39 & q2a & m0m\\
J013329.48+303614.3&  3.760& 21.520& 0.102 & 0.000 & 0.000 & 20.893& 0.125 & 0.000 & 0.000 & 21.41 & 20.56 & m1w & m0j\\
J013312.30+302355.7&  4.053& 21.457& 0.072 & 0.089 & 0.041 & 20.787& 0.062 & 0.000 & 0.000 & 21.87 & 21.33 & q2a & m0m\\
J013322.10+303731.9&  4.584& 20.768& 0.087 & 0.000 & 0.000 & 20.520& 0.082 & 0.000 & 0.000 & 21.72 & 21.03 & h5a & m0j\\
J013428.61+304820.7&  4.634& 22.243& 0.059 & 0.000 & 0.000 & 21.389& 0.085 & 0.000 & 0.000 & 21.89 & 20.45 & b4w & m03\\
J013350.94+303117.4&  4.767& 21.559& 0.060 & 0.158 & 0.015 & 20.927& 0.035 & 0.564 & 0.025 & 21.33 & 20.46 & jbw & m0e\\
J013320.60+303458.4&  4.889& 21.090& 0.137 & 0.123 & 0.024 & 21.004& 0.128 & 0.000 & 0.000 & 21.53 & 20.85 & m3w & m0k\\
J013319.10+303305.6&  4.933& 21.445& 0.048 & 0.000 & 0.000 & 20.931& 0.068 & 0.000 & 0.000 & 21.30 & 20.78 & m4w & m0k\\
J013307.46+302512.6&  5.166& 21.949& 0.037 & 0.000 & 0.000 & 21.250& 0.033 & 0.000 & 0.000 & 21.90 & 21.37 & q2a & m0m\\
J013328.77+303753.4&  5.176& 21.196& 0.054 & 0.163 & 0.016 & 20.954& 0.065 & 0.364 & 0.050 & 21.29 & 20.40 & m1w & m0j\\
J013342.17+304841.0&  5.177& 21.321& 0.102 & 0.040 & 0.005 & 20.961& 0.047 & 0.063 & 0.003 & 21.76 & 21.14 & d2a & m0b\\
J013341.39+304736.4&  5.190& 21.837& 0.051 & 0.148 & 0.013 & 21.152& 0.049 & 0.152 & 0.009 & 21.69 & 21.05 & d2a & m0b\\
J013316.08+302051.8&  5.204& 21.642& 0.065 & 0.000 & 0.000 & 21.055& 0.058 & 0.000 & 0.000 & 21.48 & 20.91 & q1a & m0n\\
J013405.98+303454.0&  5.281& 21.034& 0.048 & 0.047 & 0.011 & 20.539& 0.024 & 0.000 & 0.000 & 21.02 & 20.38 & j2w & m0e\\
J013321.78+303234.9&  5.282& 21.685& 0.298 & 0.398 & 0.119 & 20.858& 0.038 & 0.000 & 0.000 & 21.31 & 20.78 & m4w & m0k\\
J013308.37+302607.0&  5.314& 21.774& 0.057 & 0.000 & 0.000 & 21.151& 0.055 & 0.000 & 0.000 & 22.07 & 21.39 & n3a & m0l\\
J013408.64+303754.7&  5.316& 21.287& 0.058 & 0.000 & 0.000 & 20.693& 0.089 & 0.110 & 0.021 & 21.08 & 20.20 & g2w & m0d\\
J013325.71+302138.7&  5.571& 21.538& 0.045 & 0.076 & 0.007 & 21.272& 0.029 & 0.128 & 0.010 & 21.54 & 21.01 & q1a & m0m\\
J013321.51+302132.0&  5.683& 21.555& 0.066 & 0.074 & 0.015 & 20.974& 0.037 & 0.092 & 0.018 & 21.54 & 21.00 & q1a & m0m\\
J013332.14+303002.3&  5.791& 21.803& 0.108 & 0.207 & 0.015 & 20.981& 0.030 & 0.067 & 0.004 & 24.53 & 23.07 & m6a & m0f\\
J013429.09+303735.1&  5.828& 21.782& 0.034 & 0.157 & 0.007 & 21.584& 0.074 & 0.170 & 0.014 & 21.81 & 21.14 & f1a & m05\\
J013425.11+303541.4&  5.868& 20.809& 0.033 & 0.000 & 0.000 & 20.667& 0.081 & 0.000 & 0.000 & 21.80 & 21.12 & f1a & m05\\
J013350.99+303156.4&  5.889& 21.403& 0.091 & 0.254 & 0.061 & 20.762& 0.113 & 0.417 & 0.146 & 21.14 & 20.25 & jcw & m0e\\
J013407.94+303831.2&  5.902& 21.060& 0.028 & 0.000 & 0.000 & 20.796& 0.072 & 0.144 & 0.024 & 21.11 & 20.24 & gdw & m0d\\
J013359.89+303800.0&  5.997& 21.204& 0.057 & 0.000 & 0.000 & 20.698& 0.065 & 0.000 & 0.000 & 20.52 & 19.83 & gew & m0d\\
J013349.55+304743.7&  6.010& 21.718& 0.031 & 0.000 & 0.000 & 21.161& 0.078 & 0.000 & 0.000 & 21.39 & 20.53 & d4w & m0b\\
J013333.08+304230.8&  6.015& 21.802& 0.065 & 0.000 & 0.000 & 21.165& 0.053 & 0.000 & 0.000 & 21.10 & 20.55 & h1w & m0c\\
J013350.67+303445.9&  6.031& 21.063& 0.064 & 0.235 & 0.016 & 21.013& 0.034 & 0.612 & 0.021 & 21.20 & 20.51 & j2a & m0e\\
J013325.62+303510.8&  6.111& 21.010& 0.040 & 0.000 & 0.000 & 20.659& 0.084 & 0.000 & 0.000 & 21.41 & 20.71 & m3w & m0k\\
J013328.37+303730.9&  6.116& 21.238& 0.057 & 0.000 & 0.000 & 20.621& 0.069 & 0.000 & 0.000 & 21.30 & 20.39 & m1w & m0j\\
J013353.70+303152.0&  6.276& 21.709& 0.103 & 0.491 & 0.048 & 20.788& 0.056 & 0.041 & 0.003 & 21.53 & 20.92 & j4a & m0e\\
J013346.45+304740.5&  6.333& 21.272& 0.065 & 0.100 & 0.015 & 20.557& 0.085 & 0.000 & 0.000 & 21.31 & 20.45 & d4w & m0b\\
J013401.87+304148.3&  6.333& 21.714& 0.035 & 0.000 & 0.000 & 21.128& 0.071 & 0.369 & 0.033 & 21.07 & 20.16 & g1w & m0c\\
J013400.58+303630.6&  6.346& 20.846& 0.021 & 0.000 & 0.000 & 20.335& 0.048 & 0.000 & 0.000 & 20.68 & 20.06 & gew & m0d\\
J013351.82+303310.9&  6.504& 22.100& 0.093 & 0.000 & 0.000 & 21.279& 0.041 & 0.417 & 0.175 & 21.19 & 20.35 & jaw & m0e\\
J013329.22+303136.0&  6.530& 21.020& 0.049 & 0.000 & 0.000 & 20.496& 0.042 & 0.000 & 0.000 & 22.88 & 22.26 & m6a & m0k\\
J013315.71+303319.3&  6.557& 21.464& 0.161 & 0.084 & 0.019 & 20.615& 0.052 & 0.084 & 0.015 & 21.45 & 20.96 & m4w & m0k
\enddata
\end{deluxetable*}

\begin{deluxetable*}{lrrrrrrr}
\tablecaption{Stars found within 2$\arcsec$ of Cepheid variables (Abridged)\label{tb:comp}}
\tablehead{
\colhead{ID} & \colhead{$\Delta x$} & \colhead{$\Delta y$} & \colhead{$\Delta r$} & \colhead{V}     & \colhead{$\sigma_V$} & \colhead{I}     & \colhead{$\sigma_I$}\\
\colhead{}   &                \multicolumn{3}{c}{($\arcsec$)}                     &                                \multicolumn{4}{c}{(mag)}}
\startdata
J013302.03+302553.4-c001& -0.112& 0.010& 0.112& 24.021& 0.434&  \nd  &  \nd \\
J013302.03+302553.4-c002& -0.109& 0.221& 0.247& 25.333& 0.134& 25.139& 0.115\\
J013302.03+302553.4-c003&  0.333&-0.132& 0.358& 26.490& 0.091&  \nd  &  \nd \\
J013302.03+302553.4-c004& -0.306& 0.260& 0.402& 26.442& 0.190& 25.226& 0.338\\
J013302.03+302553.4-c005&  0.375& 0.203& 0.427& 25.891& 0.126& 24.051& 0.085\\
J013302.03+302553.4-c006&  0.454&-0.033& 0.455& 25.141& 0.119& 24.446& 0.099\\
J013302.03+302553.4-c007&  0.047& 0.567& 0.569& 27.163& 0.113&  \nd  &  \nd \\
J013302.03+302553.4-c008&  0.128& 0.652& 0.664& 27.007& 0.081&  \nd  &  \nd \\
J013302.03+302553.4-c009&  0.109&-0.743& 0.751& 26.728& 0.069& 26.302& 0.096\\
J013302.03+302553.4-c010& -0.790&-0.128& 0.800& 22.802& 0.071& 21.326& 0.048\\
J013302.03+302553.4-c011& -0.787&-0.223& 0.818& 25.424& 0.247&  \nd  &  \nd \\
J013302.03+302553.4-c012&  0.069& 0.870& 0.872& 27.152& 0.145&  \nd  &  \nd \\
J013302.03+302553.4-c013&  0.877& 0.051& 0.878& 27.355& 0.143&  \nd  &  \nd \\
J013302.03+302553.4-c014&  0.855&-0.279& 0.899& 26.387& 0.133& 25.761& 0.105\\
J013302.03+302553.4-c015& -0.853& 0.323& 0.912& 27.702& 0.143&  \nd  &  \nd \\
J013302.03+302553.4-c016&  0.628& 0.738& 0.969& 25.763& 0.101& 25.536& 0.100\\
J013302.03+302553.4-c017&  0.289&-0.941& 0.985& 26.810& 0.114& 26.077& 0.119\\
J013302.03+302553.4-c018& -1.016& 0.074& 1.019& 24.672& 0.065& 23.880& 0.071\\
J013302.03+302553.4-c019&  0.588& 0.858& 1.040& 27.127& 0.113&  \nd  &  \nd \\
J013302.03+302553.4-c020& -0.842& 0.663& 1.071& 26.431& 0.096& 26.194& 0.111\\
J013302.03+302553.4-c021&  0.512&-0.966& 1.093& 27.062& 0.111& 26.481& 0.094
\enddata
\end{deluxetable*}

\vspace{-24pt}

\vspace{-24pt}

\clearpage

\begin{deluxetable}{llrrr}
\tabletypesize{\footnotesize}
\tablewidth{0pc}
\tablecaption{Blending statistics \label{tb:blendstat}}
\scriptsize
\tablehead{
\colhead{Blending}  & \colhead{Sub-}   & N & \multicolumn{2}{c}{Blending criteria}\\
\colhead{level}     & \colhead{sample} &   & \multicolumn{1}{c}{CMP12} & \multicolumn{1}{c}{M01} \\
\colhead{}          & \colhead{}       &   & \colhead{\%} & \colhead{\%}}
\startdata
$S_V=0$   & all             & 149 &  $55\pm4$ &  $30\pm4$  \\
$S_V=0$   & $P<10$d         &  72 &  $50\pm6$ &  $22\pm5$  \\
$S_V=0$   & $P>10$d         &  77 &  $57\pm6$ &  $35\pm5$  \\
$S_V=0$   & $\Sigma_V<21.4$ &  71 &  $61\pm6$ &  $30\pm5$  \\
$S_V=0$   & $\Sigma_V>21.4$ &  78 &  $47\pm6$ & \B $28\pm5$ \\
\hline
$S_V<0.1$ & all             & 149 &  $73\pm4$ & \T $45\pm4$ \\
$S_V<0.1$ & $P<10$d         &  72 &  $74\pm5$ &  $43\pm6$  \\
$S_V<0.1$ & $P>10$d         &  77 &  $70\pm5$ &  $46\pm6$  \\
$S_V<0.1$ & $\Sigma_V<21.4$ &  71 &  $76\pm5$ &  $44\pm6$  \\
$S_V<0.1$ & $\Sigma_V>21.4$ &  78 &  $68\pm5$ & \B $45\pm6$  \\
\hline                                                      
$S_I=0$   & all             & 149 &  $60\pm4$ & \T $30\pm4$  \\
$S_I=0$   & $P<10$d         &  72 &  $56\pm6$ &  $19\pm4$  \\
$S_I=0$   & $P>10$d         &  77 &  $61\pm6$ &  $38\pm6$  \\
$S_I=0$   & $\Sigma_I<20.7$ &  74 &  $60\pm6$ &  $24\pm5$  \\
$S_I=0$   & $\Sigma_I>20.7$ &  75 &  $57\pm6$ & \B $33\pm5$  \\
\hline                                                      
$S_I<0.1$ & all             & 149 &  $72\pm4$ & \T $41\pm4$  \\
$S_I<0.1$ & $P<10$d         &  72 &  $71\pm5$ &  $33\pm5$  \\
$S_I<0.1$ & $P>10$d         &  77 &  $71\pm5$ &  $47\pm5$  \\
$S_I<0.1$ & $\Sigma_I<20.7$ &  74 &  $66\pm5$ &  $32\pm5$  \\
$S_I<0.1$ & $\Sigma_I>20.7$ &  75 &  $76\pm5$ &  $48\pm5$  
\enddata
\tablecomments{CMP12: this work; M01: \citet{mochejska01}}
\end{deluxetable}

\begin{figure}[t]
\includegraphics[width=0.49\textwidth]{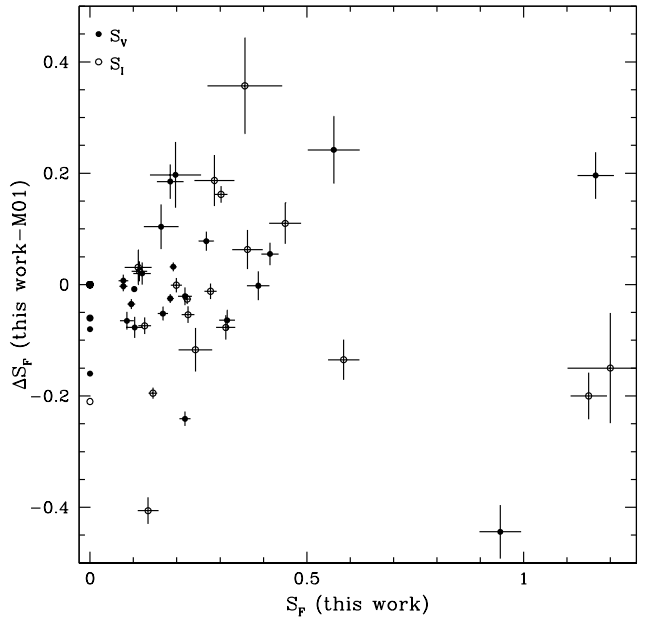}
\caption{Comparison of blending values for Cepheids in common (found in WFPC2 images) with \citet{mochejska01}; our analysis was redone using their criteria. Filled (open) circles are used to plot the blending values in the V (I) filter.}
\label{fig:compm01}
\end{figure}

\noindent{a companion flux cutoff of 6\%. The results are tabulated in the rightmost column of Table 4. We also compared the individual blending values measured for 33 variables in common in $V$ and 28 in $I$. As seen in Figure~\ref{fig:compm01}, there is good agreement with $\langle\Delta S_F\rangle=-0.02\pm0.13$.}

The statistics derived using the criteria of \citet{mochejska01} are in excellent agreement with the values presented in their paper. For example, the fraction of Cepheids with $S_V<0.1$ becomes $45\pm4$\%, compared to their value of $\sim43\pm5$\% (inferred from their Fig.~4 and Table 2). We also obtain identical values for the mean and median blending levels ($24\pm3$\% and 13\%) and reproduce the difference in blending statistics for ``short'' and ``long'' period Cepheids. Clearly, the differences between the two sets of values presented in Table~4 are due to the $2\times$ smaller critical radius adopted in our study, and emphasize the importance of angular resolution.

We also compared the disk surface brightness values we derived with those determined by \citep{mochejska01} and found agreement at the level of 0.2~\msa. We note that our surface brightness calculation is based on the average background level of the HST images within $7\arcsec$ of the Cepheid, while \citet{mochejska01} used the DIRECT ground-based images to calculate the sky in an annulus about $6\arcsec$ from the Cepheid. Regardless of the method used to measure the background or the blending criteria adopted, there is little (if any) correlation between blending fraction and surface brightness for the range of values considered here.

\citet{bresolin05} calculated blending statistics for a small sample of 16 Cepheids in NGC$\,$300 using HST/ACS images. They found a median value of 0\% and an average value of 7\%. Our results are consistent with their findings.

\section{Concluding Remarks}\label{sec:conclusions}

We have presented a survey of Cepheids in M33 and their companions within 2$\arcsec$, as resolved by HST with the ACS and WFPC2 cameras.  We calculated the flux contribution of the companions when they are blended (unresolved) in ground-based images with a seeing of $0\farcs75$. We find that more than half of the Cepheids in our sample exhibit no blending at $V$ and $I$, regardless of period or surface brightness. The majority of companion stars are located in the red giant branch and do not significantly alter the derived color of the Cepheids.

\clearpage

We plan to combine the ground-based photometry of \citet{pellerin11} with the blending values derived in this paper to investigate possible biases in the determination of distance moduli and ``metallicity corrections'' when using samples that lack such higher-resolution imaging. Additionally, our compilation of companions may be useful to derive empirical photometric bias corrections for Cepheids in more distant galaxies studied with the {\it Hubble Space Telescope}, provided the variables are located in similar environments to the M33 sample.

\vspace{-9pt}

\acknowledgements
JMC acknowledges support by the Department of Education through the GAANN Fellowship Program. AP and LMM acknowledge financial support through a Texas A\&M University faculty startup fund. We thank Profs. Jianhua Huang and Lan Zhou for useful discussions on statistical techniques and the referee, Dr. Barry Madore, for his very helpful comments.

\vspace{4pt}

Based on observations made with the NASA/ESA Hubble Space Telescope and obtained using the Mikulski Archive for Space Telescopes and the Hubble Legacy Archive (HLA) at STScI. STScI is operated by the Association of Universities for Research in Astronomy, Inc. under NASA contract NAS 5-26555. The HLA is a collaboration between STScI/NASA, the Space Telescope European Coordinating Facility (ST-ECF/ESA) and the Canadian Astronomy Data Centre (CADC/NRC/CSA).

\vspace{4pt}

{\it Facilities:} \facility{HST (WFPC2, ACS)}, \facility{WIYN (MiniMo)}   
\bibliographystyle{apj}
\bibliography{blending}

\end{document}